\documentstyle[11pt]{article}

\makeatletter
\@addtoreset{equation}{section}
\makeatother

\def\dalemb#1#2{{\vbox{\hrule height .#2pt
        \hbox{\vrule width.#2pt height#1pt \kern#1pt
                \vrule width.#2pt}
        \hrule height.#2pt}}}

\def\gtlt{\mathrel{\raise4.5pt\hbox{\oalign{$\scriptstyle>$\crcr
$\scriptstyle<$}}}}

\newcommand{\be}{\begin{equation}}
\newcommand{\ee}{\end{equation}}
\newcommand{\bea}{\begin{eqnarray}}
\newcommand{\eea}{\end{eqnarray}}

\def\ft#1#2{{\textstyle{{\scriptstyle #1}\over {\scriptstyle #2}}}}
\def\fft#1#2{{#1 \over #2}}

\newcommand{\half}{{1\over2}}

\newcommand{\auth}{James T.~Liu and H.~Sati}

\begin{document}
\begin{flushright}
\hfill{UM-TH-00-22} \\
\hfill{\bf hep-th/0009184}\\
\end{flushright}

\vspace{2cm}

\begin{center}

{\large {\bf Breathing mode compactifications and supersymmetry of the
brane-world}}

\vspace{1cm}

\auth

\vspace{10pt}
{\it
Randall Laboratory, Department of Physics, University of Michigan,\\
Ann Arbor, MI 48109--1120}
\vspace{40pt}

\underline{ABSTRACT}
\end{center}

\vspace{20pt}

It has recently been shown that the Randall-Sundrum brane-world may be
obtained from an appropriate doubled D3-brane configuration in type IIB
theory.  This corresponds, in five dimensions, to a sphere compactification
of the original IIB theory with a non-trivial breathing mode supporting the
brane.  In this paper, we shall study the supersymmetry of this reduction to
massive five-dimensional supergravity, and derive the effective
supersymmetry transformations for the fermionic superpartners to the
breathing mode.  We also consider the sphere compactifications of
eleven-dimensional supergravity to both four and seven dimensions.  For
the compactifications on $S^5$ and $S^7$, we include a squashing mode
scalar and discuss the truncation from $N=8$ to $N=2$ supersymmetry.

\pagebreak
\setcounter{page}{1}


\section{Introduction}

The Randall-Sundrum brane-world idea has attracted much recent attention
as a mechanism for trapping gravity with an uncompact extra dimension.
Although this idea leads to interesting new phenomenology and cosmology,
for a while it was not clear how fundamental this idea was, as it did
not seem to fit into any established framework for quantum gravity.
Along these lines, even the supersymmetry of the scenario itself was
in doubt.  In particular, general no-go theorems were proven
indicating that Randall-Sundrum type domain wall solutions cannot be
obtained in the context of $D=5$, $N=2$ supergravity
\cite{Kallosh:2000tj,BeCvII}.  Additionally, it was shown in
\cite{Kraus:1999it} that the Randall-Sundrum brane tension is greater
than that expected from a corresponding stack of D3-branes, thus
hinting that the brane-world is necessarily non-BPS.

Despite these difficulties with supersymmetry, it was suggested in
\cite{Verlinde,Gubser,Nojiri:2000eb,Giddings,Hawking} that the warped
geometry of the brane-world lends itself to a natural interpretation in
the framework of Maldacena's AdS/CFT correspondence.  In this context,
the Randall-Sundrum brane may be viewed as a boundary imposed on a
given horospherical slice of AdS$_5$.  Since this is no longer a
boundary at infinity, gravity no longer decouples, and the
correspondence is thus one between AdS$_5$ with a boundary and a cutoff
CFT coupled to gravity.  This provides an alternate explanation for the
trapping of gravity by the brane.

The picture of the Randall-Sundrum mechanism as a generalized AdS/CFT
conjecture was given further support in \cite{Duffliu}, where the
one-loop corrected graviton propagator calculated in the $N=4$ SCFT
was shown to agree with the classical graviton propagator obtained for
the Randall-Sundrum AdS geometry.  While not a direct test of the
supersymmetry of the brane-world, it nevertheless provided further
evidence for at least compatibility of the brane-world with supersymmetry.

Despite the tension question of \cite{Kraus:1999it}, progress has also been
made in directly relating the Randall-Sundrum model to compactifications of
ten-dimensional type IIB theory in the presence of D3-branes
\cite{Cvetic:1999fe,Cvetic:2000dz,deAlwis:2000qc,deAlwis:2000pr}.  These
compactifications have the feature of retaining the breathing mode scalar
\cite{Bremer}, to be used in supporting the five-dimensional domain wall
solution.  It has subsequently been realized in \cite{Duffliustelle}
that it is precisely the breathing mode scalar that allows the evasion
of the supergravity no-go theorems mentioned above.  In particular,
since the breathing mode is a massive mode, it lies outside the context
of \cite{Kallosh:2000tj,BeCvII}, which only considers scalars in
massless $N=2$ vector multiples.

Additionally, the realization of the
brane-world in \cite{Duffliustelle} corresponds to a doubled D3-brane
geometry in ten dimensions, where the Randall-Sundrum brane is in fact a
composite of a negative tension D3-brane and an orbifold plane.  This
structure of the Randall-Sundrum brane both stabilizes the negative
tension D3-brane and provides an exact matching of ten- and
five-dimensional tensions, thus eliminating the objection of
Ref.~\cite{Kraus:1999it} towards supersymmetry of the overall scenario.
As a result, it is now understood that the brane-world is in fact
supersymmetric, and has a natural D3-brane realization in the context of
ten-dimensional type IIB theory.

Because of the importance of the breathing mode scalar to supersymmetric
realizations of the brane-world, such massive supergravity models are
worth further investigation.  In fact, the discussion of
\cite{Duffliustelle} assumes a standard superpotential and supersymmetry
transformations for the breathing mode.  While this is clearly valid on
general $D=5$, $N=2$ grounds, it is nevertheless enlightening to obtain
the supersymmetry of the breathing mode directly from the reduction
itself.  Ref.~\cite{Bremer} has provided the general basis for sphere
reductions of supergravities with a breathing mode (and possibly an
additional squashing mode as well).  In this paper, we extend the
results of \cite{Bremer} by reducing the fermionic supersymmetry
transformations on the sphere to obtain the corresponding
lower-dimensional supersymmetries.  These transformations are necessary
for the construction of Killing spinors in the presence of the
breathing mode, and hence play a key role in understanding the
supersymmetry of the brane-world.

The squashing modes considered in \cite{Bremer} correspond to distorting
along a $U(1)$ fiber in the cases where the odd-dimensional spheres
$S^{2n+1}$ may be written as a $U(1)$ bundle over $CP^n$.  This also may be
seen as breaking the $SO(2n+2)$ isometry group (which is also the $R$-symmetry
group) down to $SU(n+1)\times U(1)$.  For both the $S^5$ compactification of
type IIB theory and the $S^7$ compactification of $D=11$ supergravity, this
squashing corresponds to the breaking of $N=8$ to $N=2$ supersymmetry
through the retention of $SU(n+1)$ singlets only.  What makes the fermion
supersymmetries non-trivial in these $U(1)$ fibration cases is the fact
that the $N=2$ supercharges are charged under this $U(1)$.  Since this charge
corresponds to Kaluza-Klein momentum in the fiber direction, most previous
investigations of such fibered spaces have further truncated to the $U(1)$
neutral sector \cite{Nilsson,Duff:1997qz,Duff55} at the expense of losing
some or all supersymmetry.  For example, in the $S^7$ case, the
neutral Killing spinors give rise to either $N=6$ or $N=0$ supersymmetry,
and not the maximal $N=8$ of the complete (untruncated) $S^7$
compactification \cite{Nilsson,Duff:1997qz}.  The situation is even more
drastic for the $S^5$ compactification, as there are no fermions at all
in the uncharged sector since $CP^2$ does not admit a spin structure
\cite{Hawking:1978ab,Pope:1980ub,Duff55}.

In this paper, we will take a closer look at the supersymmetry and
construction of Killing spinors for such Hopf fibered spheres.  By allowing
momentum in the fiber direction, and hence $U(1)$ charge, we are able to
demonstrate explicitly the full $N=8$ supersymmetry of the $S^5$ and $S^7$
compactifications when written in terms of $U(1)$ bundles over $CP^n$.
Keeping $U(1)$ charged spinors, we then truncate to appropriate $N=2$
limits by restricting to only singlets on $CP^n$.  We conjecture that
such a truncation to the $N=2$ breathing/squashing multiplet coupled
to $N=2$ supergravity could in fact be a consistent truncation of the
full Kaluza-Klein spectrum of the sphere compactification.

Some general features of scalars coupled to $N=2$ supergravity may be
obtained by considering a bosonic Lagrangian of the form
\begin{equation}
\label{eq:n=2lag}
e^{-1}{\cal L}_D=R-\ft12|\partial\vec\phi|^2-V(\vec\phi),
\end{equation}
with expected supersymmetry transformations
\begin{eqnarray}
\label{eq:n=2}
\delta\psi_\mu&=&\left[\nabla_\mu-\ft1{2(D-2)\sqrt{2}}W\gamma_\mu\right]
\varepsilon,\nonumber\\
\delta\vec\lambda&=&\ft12\left[\gamma\cdot\partial\vec\phi
+\sqrt{2}\partial_{\vec\phi}W\right]\varepsilon.
\end{eqnarray}
The potential $V$ may be obtained from the superpotential through the
relation
\begin{equation}
\label{eq:psp}
V=\left|\partial_{\vec\phi}W\right|^2-\fft{(D-1)}{2(D-2)}W^2.
\end{equation}
This is an example of a Ward-like identity for the scalars in a
supergravity theory \cite{cecotti}.
Of course, this $N=2$ framework is somewhat heuristic without a complete
specification of the supermultiplets ({\it e.g.}~vector, tensor or
hyper), fields and transformations.  Nevertheless, we will verify that
the $N=2$ truncations derived below satisfy the above relations.

In the following section, we examine the $S^5$ reduction of
ten-dimensional type IIB supergravity, both with and without the
inclusion of a squashing mode.  For the squashing mode, generated by
fibering $U(1)$ over $CP^2$, we discuss generalized spin structures and
construct appropriately charged Killing spinors.  Then in sections
3 and 4 we consider the $S^7$ and $S^4$ compactifications of eleven
dimensional supergravity.  For the former we also include a squashing
mode, while for the latter this is not possible.  Finally, we conclude
with some discussion on the implications for supersymmetry of the
brane-world scenario.


\section{$S^5$ Reduction of $D=10$ type IIB supergravity}

Since the five-dimensional case is of immediate attention for the
brane-world, we first consider the $S^5$ compactification of type IIB
supergravity.  Even with inclusion of the breathing mode, the
sphere reduction proceeds via a Freund-Rubin ansatz for the self-dual
5-form.  With the 5-form field-strength active, the type IIB supergravity
does not admit a covariant Lagrangian formulation.  Thus we instead
work with the equations of motion.  With only the metric and 5-form
active, the relevant bosonic equations of motion are:
\begin{eqnarray}
\label{eq:iibeom}
\hat R_{MN}&=&\fft12\fft1{2\cdot4!}F_{MPQRS}F_N{}^{PQRS},\nonumber\\
dF_{[5]}&=&0,\nonumber\\
F_{[5]}&=&*F_{[5]}.
\end{eqnarray}
In addition, the supersymmetry variation of the gravitino is given (on
this background) by:
\begin{equation}
\label{eq:10gravi}
\delta \hat\psi_M = \left[\hat\nabla_M +{i\over16\cdot5!}
F_{NPQRS}\Gamma^{NPQRS}\Gamma_M \right]\hat\epsilon,
\end{equation}
where the type IIB spinors are chiral,
\begin{equation}
\label{eq:10chiral}
\Gamma_{11}\hat\epsilon = \hat\epsilon, \qquad \Gamma_{11}\hat\psi_M =
\hat\psi_M.
\end{equation}
The transformation of the ten-dimensional dilatino vanishes trivially
for this subset of fields, and hence may be ignored.
Note that explicit ten-dimensional quantities are denoted with a
caret to avoid confusion with their five-dimensional counterparts.

\subsection{The breathing mode reduction}

The breathing mode reduction proceeds along the lines of the
Kaluza-Klein ansatz of Ref.~\cite{Bremer}:
\bea
\label{eq:kka}
ds_{10}^2&=&e^{2\alpha\varphi}ds_5^2+e^{2\beta\varphi}ds^2(S^5),\nonumber\\
F_{[5]}&=&4me^{8\alpha\varphi}\epsilon_{[5]}+4m\epsilon_{[5]}(S^5),
\eea
where
\be
\alpha=\fft14\sqrt{\fft53},\qquad\beta=-\fft35\alpha.
\ee
The resulting five-dimensional theory is described by a Lagrangian of the
form
\begin{equation}
e^{-1}{\cal L}_5=R-\ft12(\partial\varphi)^2-V(\varphi),
\end{equation}
where $V(\varphi)$ has the double exponential form
\begin{equation}
\label{eq:5pot}
V=8m^2e^{8\alpha\varphi}-R_5e^{\fft{16}5\alpha\varphi}.
\end{equation}
$R_5$ is the Ricci scalar of $S^5$, and may be viewed as a parameter of
the compactification.  This potential has an AdS$_5$ minimum at
\begin{equation}
e^{\fft{24}5\alpha\varphi_*}=\fft{R_5}{20m^2}.
\end{equation}

We now wish to reduce the $D=10$ gravitino variation, (\ref{eq:10gravi}),
to obtain the corresponding variations in the $S^5$ reduced theory.  To
carry out this reduction, we perform a convenient $5+5$ decomposition of
the ten-dimensional Dirac matrices (in tangent space):
\be
\Gamma^{\overline{M}}=(\gamma^{\overline{\mu}} \otimes 1  \otimes \sigma^1,
  1 \otimes \tilde\gamma^{\overline{a}} \otimes \sigma^2)
\ee
where
\bea
\{\gamma^{\overline{\mu}} ,\gamma^{\overline{\nu}}\}
&=&2\eta^{\overline{\mu}\overline{\nu}},\nonumber\\
\{\tilde\gamma^{\overline{a}},\tilde\gamma^{\overline{b}} \}
&=&2\delta^{\overline{a}\overline{b}}.
\eea
Here $\mu,\nu,\ldots=0,1,2,3,4$ are spacetime indices with
$\gamma^{4}\equiv i\gamma^{0}\gamma^{1}\gamma^{2}\gamma^{3}$
and $a,b,\ldots=5,6,7,8,9$ are internal indices with
$\tilde\gamma^{9}\equiv\tilde\gamma^{5}\tilde\gamma^{6}\tilde\gamma^{7}
\tilde\gamma^{8}$.  The ten-dimensional chirality operator is
$\Gamma^{11}=\Gamma^{0}\ldots\Gamma^{9} =1\otimes 1\otimes \sigma^3$, so
that type IIB spinors satisfying (\ref{eq:10chiral}) may be written as,
{\it e.g.}
\begin{equation}
\label{eq:10weyl}
\hat\epsilon = \varepsilon\otimes\eta\otimes\left[1\atop 0\right].
\end{equation}

For the breathing mode reduction, (\ref{eq:kka}), we compute
\bea
F_{MNPQR}\Gamma^{MNPQR}
&=&4\cdot5!m e^{3\alpha\varphi}[\Gamma^{01234}+\Gamma^{56789}]\nonumber\\
&=&-4i\cdot5!m e^{3\alpha\varphi} [1\otimes 1
\otimes (\sigma^1 + i\sigma^2)].
\eea
Furthermore, the ten-dimensional covariant derivative decomposes as
\bea
\hat\nabla_\mu&=&\nabla_\mu\otimes1\otimes1
+\fft\alpha2\gamma_\mu{}^\nu\partial_\nu\varphi\otimes1\otimes1,
\nonumber\\
\hat\nabla_a&=&1\otimes\tilde\nabla_a\otimes1+\fft{3i\alpha}{10}
e^{-\fft85\alpha\varphi}\gamma^\mu\partial_\mu\varphi\otimes\tilde\gamma_a
\otimes\sigma^3.
\eea
Hence from (\ref{eq:10gravi}) we obtain the gravitino transformations
\bea
\label{eq:gvar}
\delta\hat\psi_\mu&=&\left[\nabla_\mu+\fft\alpha2\gamma_\mu{}^\nu\partial_\nu
\varphi+\fft{m}4e^{4\alpha\varphi}\gamma_\mu\otimes1\otimes(1+\sigma^3)\right]
\hat\epsilon,\nonumber\\
\delta\hat\psi_a&=&\left[\tilde\nabla_a+\fft{3i\alpha}{10}
e^{-\fft85\alpha\varphi}\gamma^\mu\partial_\mu\varphi\otimes
\tilde\gamma_a\otimes\sigma^3+\fft{im}4
e^{\fft{12}5\alpha\varphi}\tilde\gamma_a\otimes(1+\sigma^3)\right]
\hat\epsilon.
\eea
Note that, for simplicity of notation, the tensor product structure in
spinor space is hidden in the terms which act only on a single spinor
subspace.  The chiral structure of the IIB theory is apparent in
(\ref{eq:gvar}), where, in agreement with (\ref{eq:10chiral}),
$\hat\psi_M$ must have $\sigma^3$ eigenvalue
$+1$ to transform properly along with a self-dual $F_{[5]}$.  Taking
this into account, and working with Weyl spinors of the form
(\ref{eq:10weyl}), the above simplifies to
\bea
\label{eq:susgr}
\delta\hat\psi_\mu&=&\left[\nabla_\mu+\fft\alpha2\gamma_\mu{}^\nu\partial_\nu
\varphi+\fft{m}2e^{4\alpha\varphi}\gamma_\mu\right]\hat\epsilon,\nonumber\\
\delta\hat\psi_a&=&\left[\tilde\nabla_a+\fft{3i\alpha}{10}
e^{-\fft85\alpha\varphi}\gamma^\mu\partial_\mu\varphi\otimes
\tilde\gamma_a+\fft{im}2 e^{\fft{12}5\alpha\varphi}\tilde\gamma_a\right]
\hat\epsilon.
\eea

For this Kaluza-Klein reduction, the ten-dimensional spinors $\hat\psi_M$
and $\hat\epsilon$ may be decomposed in terms of Killing spinors on the
sphere.  For the $n$-sphere, such spinors satisfy
\be
\label{eq:snkse}
\left[\tilde\nabla_{a} \pm {i\over2}\tilde\gamma_a
\sqrt{R_{n}\over n(n-1)}\,\right]\eta=0,
\ee
where $R_n$ is the Ricci scalar which fixes the size of the sphere.
Note that Killing spinors may be found for either sign of the second
term in (\ref{eq:snkse}), corresponding to the orientation of the sphere.
Specializing to the case at
hand, $\eta$ is a complex four-component spinor.  Thus there are four
independent (complex) Killing spinors on $S^5$, giving the expected
$D=5$, $N=8$ supersymmetry of the round-sphere compactification.

We proceed by using
(\ref{eq:snkse}) to eliminate the covariant derivative on $S^5$ in
(\ref{eq:susgr}).  As a result, we find that $\hat\psi_a$ has the
property of being a spinorial superpartner to the breathing mode $\varphi$,
while $\hat\psi_\mu$ survives as the five-dimensional gravitino
variation.  The breathing mode `dilatino' may be normalized by defining
\begin{equation}
\lambda^{(5)}_i \otimes \eta^i = -\fft{i}{3\alpha}e^{\fft{11}{10}\alpha\varphi}
\tilde\gamma^a\hat\psi_a,
\end{equation}
resulting in the transformation
\begin{equation}
\delta \lambda^{(5)}_i=\fft12
\left[\gamma\cdot\partial\varphi +\fft5{3\alpha}
\left(me^{4\alpha\varphi}\mp \sqrt{R_{5}\over20}e^{\fft85\alpha\varphi}\right)
\right]\varepsilon^{(5)}_i,
\end{equation}
where
\begin{equation}
\varepsilon^{(5)}_i\otimes\eta^i=e^{-\fft12\alpha\varphi}\hat\epsilon.
\end{equation}
Here, $i=1,2,3,4$ labels the Killing spinor (where all spinors are taken
to be Dirac), indicating the trivial $N=8$ structure.
The scaling of the supersymmetry parameter is natural in a Kaluza-Klein
context, and partially eliminates the $\partial_\nu\varphi$ term in the
reduction of $\delta\hat\psi_\mu$ in (\ref{eq:susgr}).  Similarly, the
five-dimensional gravitino takes on the shifted form
\be
\psi_{i\,\mu}^{(5)}\otimes\eta^i=e^{-\fft12 \alpha\varphi}\hat\psi_{\mu}
-\alpha\gamma_\mu\lambda^{(5)}_i\otimes\eta^i,
\ee
so that its supersymmetry transformation has the form
\be
\delta \psi_{i\,\mu}^{(5)}=\left [ \nabla_{\mu}-\fft16\left(2me^{4\alpha\varphi}
\mp5\sqrt{R_{5} \over 20}e^{\fft85\alpha\varphi}\right) \gamma_{\mu}
\right]\varepsilon^{(5)}_i.
\ee

Focusing on a particular component, say $i=1$, we can write the above
transformations in the form, (\ref{eq:n=2}), appropriate to $D=5$, $N=2$
supergravity:
\bea
\delta\psi_\mu&=&\left[\nabla_\mu-\ft1{6\sqrt{2}}
W\gamma_\mu\right]\varepsilon,
\nonumber\\
\delta\lambda&=&\ft12
\left[\gamma\cdot\partial\varphi+\sqrt{2}\partial_\varphi W \right]
\varepsilon,
\eea
where one can identify the superpotential as
\be
\label{eq:5spot}
W=\sqrt{2}\Biggl[2m e^{4\alpha\varphi} \mp 5 \sqrt{R_{5} \over 20}
e^{\frac{8}{5}\alpha \varphi}\Biggr].
\ee
Furthermore, it is easy to check that this superpotential and the potential
(\ref{eq:5pot}) are related by the five-dimensional version of the
identity (\ref{eq:psp}):
\be
V=(\partial_{\varphi}W)^2-\fft23W^2.
\ee
Note that the potential itself is unaffected by the sign ambiguity present
in (\ref{eq:5spot}).

\subsection{Squashing the five-sphere}

Note that $S^{2n+1}$ can be written as a $U(1)$ bundle over $CP^n$,
where in the present case $n=2$.  This suggests the inclusion of a
squashing mode, corresponding to a breaking of the $SO(6)$ isometry
group of the round $S^5$ to $SU(3)\times U(1)$.  This construction may
be made explicit by first reducing the type IIB theory on $S^1$, and
then reducing further from nine to five dimensions on $CP^2$.  We follow
the procedure given in \cite{Duff55,Bremer} while paying attention to the
fermion supersymmetries.

At this point, it is important to realize that $CP^{2n}$ does not admit
a spin structure.  Thus, at first sight, the reduction over $CP^2$ results
in a five-dimensional model without any fermions, and in particular
without supersymmetry \cite{Duff55}.  However we know
that this compactification admits fermions and is supersymmetric, because
in the appropriate limit it is nothing but reduction on the round $S^5$.
The resolution of this difficulty is the realization that, while $CP^2$
does not admit a spin structure, it nevertheless does admit a spin$^c$
structure \cite{Hawking:1978ab,Pope:1980ub}.  This essentially indicates
that spinors on $CP^2$ are charged under the $U(1)$ fiber.  As $U(1)$
charge corresponds to momentum along the $S^1$ direction, such states
may be considered `massive' in $D=9$, and hence are usually truncated
out in an ordinary Kaluza-Klein compactification.  However, since we are
interested in the supersymmetry of this model, we must retain the $U(1)$
charged fermions by allowing for dependence on the circle coordinate.
It is worth noting, though, that the bosonic fields may be truncated at
the massless Kaluza-Klein level since the background of interest, namely
the squashing mode solution of \cite{Bremer}, is already complete in the
$U(1)$ neutral sector.  This simplifies the situation, as otherwise a
complete reduction with all massive Kaluza-Klein states would be
considerably more involved.

Even with this momentum dependence in mind, the reduction to $D=9$
proceeds straightforwardly.  For a reduction on $S^1$, we write
\be
ds_{10}^2 = e^{2a\varphi}ds_9^2 + e^{2b\varphi}(dz+{\cal A}_M dx^M)^2,
\ee
where now $M,N$ are nine dimensional spacetime indices and $m,n$ are
$SO(1,8)$ tangent space indices.  The circle direction is given by $z$
and 9 for curved and tangent space values respectively.  The constants
$a$ and $b$ are
\begin{equation}
a=-\fft1{4\sqrt{7}},\qquad b=-7a.
\end{equation}
With only these fields and $F_{[5]}$ active, the reduction of the IIB
equations of motion, (\ref{eq:iibeom}), yield a set of nine-dimensional
equations that may be derived from the Lagrangian \cite{Bremer}
\begin{equation}
e^{-1}{\cal L}_9 = R-\ft12(\partial\varphi)^2-\ft14e^{-16a\varphi}
{\cal F}_{[2]}^2 - \ft1{48}e^{8a\varphi}F_{[4]}^2
-\ft12e^{-1}*(F_{[4]}\wedge F_{[4]}\wedge{\cal A}_{[1]}).
\end{equation}
Because of its self-dual nature, $F_{[5]}$ reduces to a single
four-form field strength, $F_{[4]}$, given by
$F_{[4]\,MNPQ}\equiv F_{[5]\,MNPQ\,z}$.

For the fermion variations, it is convenient to decompose the
ten-dimensional Dirac matrices (in tangent space) as
\be
\hat\Gamma^{\hat m} =(\Gamma^m \otimes \sigma^1, 1 \otimes \sigma^2)
\ee
with $\Gamma^8=i\Gamma^0\ldots\Gamma^7$.  For this choice, $\hat\Gamma^{11}
=1\otimes\sigma^3$, so that IIB spinors have the same form as before.
In the following expressions, whenever vielbeins and metric factors are
hidden, they are taken to be either ten- or nine-dimensional entities as
appropriate.  Thus, {\it e.g.},
\begin{equation}
\hat\Gamma_M = e^{a\varphi}\Gamma_M + e^{b\varphi}{\cal A}_M\Gamma^9,\qquad
\hat\Gamma_z = e^{b\varphi}\Gamma^9.
\end{equation}
This shifting of quantities by the $U(1)$ field is standard in
Kaluza-Klein reductions.

The gravitino variation, (\ref{eq:10gravi}), splits into both
spin-$\fft32$ and spin-$\fft12$ pieces in $D=9$.  Using the identity
\begin{equation}
F_{{\hat M}{\hat N}{\hat P}{\hat Q}{\hat R}}
{\hat \Gamma}^{{\hat M}{\hat N}{\hat P}{\hat Q}{\hat R}}
=5e^{3a\varphi}F_{[4]}\cdot\Gamma\,\Gamma^9(1-\hat\Gamma^{11}),
\end{equation}
and accounting for the chirality of $\hat\epsilon$, (\ref{eq:10chiral}),
we obtain after some manipulation
\begin{eqnarray}
\delta\psi^{(9)}_M&=&\Biggl[D_M+\fft{i}7e^{8a\varphi}\Gamma_M\partial_z
-\fft{i}{56}e^{-8a\varphi}(\Gamma_M{}^{NP}
-12\delta_M^N\Gamma^P){\cal F}_{NP}\nonumber\\
&&\kern6em
-\fft1{56\cdot4!}e^{4a\varphi}(\Gamma_M\Gamma^{NPQR}-7\Gamma^{NPQR}
\Gamma_M)F_{NPQR}\Biggr]\epsilon^{(9)},\nonumber\\
\delta\lambda^{(9)}&=&\fft12\Biggl[
\Gamma\cdot\partial\varphi-\fft{2i}{7a}e^{8a\varphi}\partial_z
+\fft{i}{28a}e^{-8a\varphi}{\cal F}_{[2]}\cdot\Gamma
+\fft1{28\cdot4!a}e^{4a\varphi}F_{[4]}\cdot\Gamma\Biggr]\epsilon^{(9)},
\end{eqnarray}
where the shifted quantities are defined as
\begin{eqnarray}
\psi^{(9)}_M&=&e^{-\fft12a\varphi}(\hat\psi_M-{\cal A}_M\hat\psi_z)
+ia\Gamma_M\Gamma^9\lambda^{(9)},\nonumber\\
\lambda^{(9)}&=&-\fft{i}{7a}e^{\fft{15}2a\varphi}\hat\psi_z,\nonumber\\
\epsilon^{(9)}&=&e^{-\fft12a\varphi}\hat\epsilon.
\end{eqnarray}
Here, $D_M=\nabla_M-{\cal A}_M\partial_z$ is the $U(1)$ covariant
derivative for charged spinors.  This modification, as well as the
inclusion of terms proportional to $\partial_z$ in the above, fully
accounts for the possible momentum dependence in the $z$ direction.
Other than for this $z$ dependence, after some rearrangement, these
expressions agree with the transformations derived in \cite{khvi}.

With the above transformations out of the way, we now proceed to five
dimensions using the ansatz \cite{Bremer}
\bea
ds_{9}^2&=&e^{2\alpha f}ds_5^2+e^{2\beta f}ds^2(CP^2),\nonumber\\
F_{[4]}&=&4m\epsilon_{[4]}(CP^2),\nonumber\\
{\cal F}_{[2]}&=&2\mu J_{[2]}(CP^2),
\eea
where
\be
\alpha=\sqrt{\frac{2}{21}},\qquad \beta=-\fft34 \alpha,
\ee
and $J_{[2]}$ is the Kahler form satisfying $J_{ac}J^c{}_b=-g_{ab}$ with
$g_{ab}$ the standard metric on $CP^2$.  The reduced bosonic Lagrangian
has the form
\begin{equation}
e^{-1}{\cal L}_5=R-\ft12(\partial\varphi)^2-\ft12(\partial f)^2
-V(\varphi,f),
\end{equation}
with
\begin{equation}
\label{eq:pfpot}
V(\varphi,f)=8m^2e^{8a\varphi+8\alpha f}+4\mu^2e^{-16a\varphi+5\alpha f}
-R_4e^{\fft72\alpha f}.
\end{equation}
Note that we have corrected a factor of four in the second term of $V$.

The $D=9$ Dirac matrices may now be split into space-time and $CP^2$
components
\be
\Gamma^M =(\gamma^\mu \otimes \tilde\gamma^5, 1 \otimes \tilde\gamma^a),
\ee
where the chirality operator on $CP^2$ is $\tilde\gamma^{5} =
\tilde\gamma^1 \tilde\gamma^2 \tilde\gamma^3 \tilde\gamma^4$.
Then, as in the round $S^5$ compactification, we obtain the decomposition
\begin{eqnarray}
D_\mu^{(9)}&=&\nabla_\mu\otimes1+\fft\alpha2\gamma_\mu{}^\nu\partial_\nu
f\otimes1,\nonumber\\
D_a^{(9)}&=&1\otimes\tilde D_a-\fft{3\alpha}8e^{-\fft74\alpha f}
\gamma\cdot\partial f\otimes\tilde\gamma_a\tilde\gamma^5.
\end{eqnarray}
We also make use of the identities
\begin{equation}
F_{[4]}\cdot\Gamma=4\cdot4!me^{3\alpha f}\tilde\gamma^5
\end{equation}
and
\begin{equation}
{\cal F}_{[2]}\cdot\Gamma=-2i\mu e^{\fft32\alpha f}\tilde Q\tilde\gamma^5,
\end{equation}
where
\begin{equation}
\label{eq:defq}
\tilde Q\equiv iJ_{[2]}\cdot\tilde\gamma\tilde\gamma^5.
\end{equation}

The relation between $U(1)$ charge and momentum in the $z$
direction can be made more precise.  Following \cite{Duff:1997qz}, we
note that the period of $z$ must satisfy
\begin{equation}
\Delta z=\int{\cal F}_{[2]}=2\mu\int J.
\end{equation}
Since $R_4$ is defined as the Ricci scalar of $CP^2$, we have
$R_{ab}=\fft{R_4}4g_{ab}$, so that $\rho = \fft{R_4}4J$ where $\rho$ is
the Ricci form.  Using $c_1=\fft1{2\pi}\int\rho$ and $c_1(CP^n)=n+1$
where $c_1$ is the first Chern class, we finally obtain
\begin{equation}
\Delta z=\fft{48\pi\mu}{R_4}.
\end{equation}
Thus, defining the circle radius $L$ by $z=z+2\pi L$, we have
$L=24\mu/R_4$.  As a result, for a mode expansion in harmonics of the
form $e^{iqz/L}$, we may replace $\partial_z$ by
\begin{equation}
\partial_z=\fft{iq}L=i\left(\fft{R_4}{24\mu}\right)q
\end{equation}
where $q$ may be considered to be the $U(1)$ charge.

Putting everything together, we find
\begin{eqnarray}
\label{eq:bs9susy}
\delta\psi_\mu^{(9)}&=&\Biggl[\nabla_\mu+\fft\alpha2\gamma_\mu{}^\nu
\partial_\nu f-\fft17\left(\fft{R_4}{24\mu}\right)q\,e^{8a\varphi+\alpha f}
\gamma_\mu\otimes\tilde\gamma^5\nonumber\\
&&\qquad\qquad-\fft{\mu}{28}e^{-8a\varphi+\fft52\alpha f}\gamma_\mu
\otimes\tilde Q+\fft{3m}7e^{4a\varphi+4\alpha f}\gamma_\mu
\Biggr]\epsilon^{(9)},\nonumber\\
\delta\psi_a^{(9)}&=&\Biggl[\tilde D_a-\fft{3\alpha}8 e^{-\fft74\alpha f}
\gamma\cdot\partial f\otimes\tilde\gamma_a\tilde\gamma^5
-\fft17\left(\fft{R_4}{24\mu}\right)q\,e^{8a\varphi-\fft34\alpha f}
\tilde\gamma_a\nonumber\\
&&\qquad\qquad-\fft{i\mu}{28}e^{-8a\varphi+\fft34\alpha f}
(\tilde\gamma_a{}^{bc}-12\delta_a^b\tilde\gamma^c)J_{bc}
-\ft{4m}7e^{4\alpha\varphi+\fft94\alpha f}
\tilde\gamma_a\tilde\gamma^5\Biggr]\epsilon^{(9)},\nonumber\\
\delta\lambda^{(9)}&=&\ft12e^{-\alpha f}\tilde\gamma^5
\Biggl[\gamma\cdot\partial\varphi
+\fft2{7a}\left(\fft{R_4}{24\mu}\right)q\,e^{8a\varphi+\alpha f}
\tilde\gamma^5\nonumber\\
&&\qquad\qquad+\fft{\mu}{14a}e^{-8a\varphi+\fft52\alpha f}\tilde Q
+\fft{m}{7a}e^{4a\varphi +4\alpha f}
\Biggr]\epsilon^{(9)}.
\end{eqnarray}
To proceed, we now need to determine the form of the Killing spinors on
the Hopf fibration of $S^5$.  This may be done by realizing that the
conventional Freund-Rubin compactification on the round $S^5$ is obtained
when the breathing and squashing modes are turned off.  This occurs when
both $\varphi$ and $f$ are sitting at the $N=8$ critical point of the
potential (\ref{eq:pfpot}), corresponding to the constant values
\begin{equation}
\mu^2=m^2e^{24a\varphi_*+3\alpha f_*},\qquad
\left(\fft{R_4}{24}\right)=m^2e^{8a\varphi_*+\fft92\alpha f_*}.
\end{equation}
For this case, the gradient terms drop out from (\ref{eq:bs9susy}), and one
finds
\begin{eqnarray}
\label{eq:ksecp}
\delta\lambda^{(9)}&=&\fft1{28a} e^{\fft34\alpha f_*} \tilde\gamma^5
\sqrt{\fft{R_4}{24}}\,(4q\tilde\gamma^5+\tilde Q\pm2)\epsilon^{(9)},\nonumber\\
\delta(\psi_a^{(9)}+ae^{-\fft34\alpha f_*}\tilde\gamma_a\lambda^{(9)})
&=&\left[\tilde D_a-\fft12\sqrt{\fft{R_4}{24}}(\pm\tilde\gamma_a
\tilde\gamma^5-iJ_{ab}\tilde\gamma^b)\right]\epsilon^{(9)}.
\end{eqnarray}
The sign ambiguity arises by considering the two cases,
\begin{equation}
\mu=\pm me^{12\alpha\varphi_*+\fft32\alpha f_*},
\end{equation}
corresponding to the choice of orientation of $S^5$.  Now, the
vanishing of $\delta\lambda^{(9)}$ then imposes the condition,
\begin{equation}
\label{eq:qQcond}
(4q\tilde\gamma^5+\tilde Q\pm2)\eta=0,
\end{equation}
on Killing spinors $\eta$.  From the definition of $\tilde Q$,
(\ref{eq:defq}), it is easy to verify that it has eigenvalues $(-4,0,0,4)$,
with corresponding $\tilde\gamma^5$ eigenvalues $(-1,1,1,-1)$.  Focusing
on the positive sign in (\ref{eq:qQcond}), we see that it
may be satisfied for $U(1)$ charges $q=(-\fft12,-\fft12,
-\fft12,\fft32)$.  Note that this verifies the decomposition of the
complex spinor representation $4\to3_{-1/2}+1_{3/2}$ under the split
$SO(6) \supset SU(3)\times U(1)$.  For the negative sign, we would
obtain instead $q=(-\fft32,\fft12,\fft12,\fft12)$, which is the charge
conjugate of the above.

It remains to consider the gravitino variation on $CP^2$.  Here it is easier
not to consider the variation (\ref{eq:ksecp}) directly, but rather to check
the integrability condition
\begin{equation}
[\tilde{\cal D}_a,\tilde{\cal D}_b] = \fft14R_{abcd}\tilde\gamma^{cd}
+\fft{R_4}{48}[-\tilde\gamma_{ab}-2iJ_{ab}(2q\pm\tilde\gamma^5)-J_{ac}J_{bd}
\tilde\gamma^{cd}].
\end{equation}
Substituting the Riemann tensor on $CP^2$,
\begin{equation}
R_{abcd}=\fft{R_4}{24}[g_{ac}g_{bd}-g_{ad}g_{bc}+J_{ac}J_{bd}
-J_{ad}J_{bc}+2J_{ab}J_{cd}],
\end{equation}
we then find
\begin{equation}
[\tilde{\cal D}_a,\tilde{\cal D}_b] = -i\fft{R_4}{48}J_{ab}\tilde\gamma^5
(4q\tilde\gamma^5+\tilde Q\pm2),
\end{equation}
which indeed vanishes for Killing spinors satisfying (\ref{eq:qQcond}).
We have thus found the expected four independent complex Killing spinors for
the $U(1)$ fibered $CP^2$ construction.

Following the procedure applied previously to the round $S^5$, we now
return to the full supersymmetry transformations, (\ref{eq:bs9susy}), and
decompose the various fermions in terms of the above set of Killing spinors.
Defining
\begin{eqnarray}
\varepsilon^{(5)}\otimes\eta&=&e^{-\fft12\alpha f}\epsilon^{(9)},\nonumber\\
\lambda^{(5)}\otimes\eta&=&e^{\fft12\alpha f}\tilde\gamma^5\lambda^{(9)},
\nonumber\\
\chi^{(5)}\otimes\eta&=&\ft1{3\alpha}e^{\fft54\alpha f}\tilde\gamma^a
\tilde\gamma^5\psi_a^{(9)},\nonumber\\
\psi_\mu^{(5)}\otimes\eta&=&e^{-\fft12\alpha f}\psi_\mu^{(9)}
-\alpha\gamma_\mu\chi^{(5)}\otimes\eta,
\end{eqnarray}
and using (\ref{eq:qQcond}) to eliminate the charge $q$,
we find the resulting set of five-dimensional supersymmetry transformations:
\begin{eqnarray}
\delta\lambda^{(5)}&=&\fft12\Bigl[\gamma\cdot\partial\varphi
-\fft1{14a}\left(\fft{R_4}{24\mu}\right)(\tilde Q\pm2)e^{8a\varphi+\alpha f}
+\fft\mu{14a}\tilde Qe^{-8a\varphi+\fft52\alpha f}
+\fft{m}{7a}e^{4a\varphi+4\alpha f}\Bigr]\varepsilon^{(5)},\nonumber\\
\delta\chi^{(5)}&=&\fft12\Bigl[\gamma\cdot\partial f
+\fft1{3\alpha}\sqrt{\fft{R_4}{24}}(\tilde Q\mp4)e^{\fft74\alpha f}
-\fft2{21\alpha}\left(\fft{R_4}{24\mu}\right)(\tilde Q\pm2)
e^{8a\varphi+\alpha f}\nonumber\\
&&\kern6em-\fft{5\mu}{21\alpha}\tilde Q e^{-8a\varphi+\fft52\alpha f}
+\fft{32m}{21\alpha}e^{4a\varphi+4\alpha f}\Bigr]\varepsilon^{(5)},
\nonumber\\
\delta\psi_\mu^{(5)}&=&\Bigl[\nabla_\mu-\fft16\sqrt{\fft{R_4}{24}}
(\tilde Q\mp4)e^{\fft74\alpha f}\gamma_\mu
+\fft1{12}\left(\fft{R_4}{24\mu}\right)(\tilde Q\pm2)e^{8a\varphi+\alpha f}
\gamma_\mu\nonumber\\
&&\kern6em+\fft\mu{12}\tilde Qe^{-8a\varphi+\fft52\alpha f}\gamma_\mu
-\fft{m}3e^{4a\varphi+4\alpha f}\gamma_\mu\Bigr]\varepsilon^{(5)}.
\end{eqnarray}
(one may consider there to be four sets of such equations---one for each of
the four eigenvalues of $\tilde Q$.)

Remarkably, these transformations follow the $N=2$ form, (\ref{eq:n=2}),
with `superpotential'
\begin{equation}
\label{eq:5sqsp}
W=\sqrt{2}\left[\sqrt{\fft{R_4}{24}}(\tilde Q\mp4)e^{\fft74\alpha f}
-\fft12\left(\fft{R_4}{24\mu}\right)(\tilde Q\pm2)e^{8a\varphi+\alpha f}
-\fft\mu2\tilde Qe^{-8a\varphi+\fft52\alpha f}
+2me^{4a\varphi+4\alpha f}\right].
\end{equation}
While this holds for all valid choices of $\tilde Q$ eigenvalue, based on
the truncation to the $N=2$ model with squashing mode, we are only
interested in retaining the $SU(3)$ singlet state $1_{3/2}$, with
$\tilde Q=4$ (or $1_{-3/2}$ with $\tilde Q=-4$).  In this case, the
superpotential reads
\begin{equation}
W=\sqrt{2}\left[2me^{4a\varphi+4\alpha f}\mp2\mu e^{-8a\varphi+\fft52\alpha f}
\mp\fft{R_4}{8\mu}e^{8a\varphi+\alpha f}\right],
\end{equation}
and it may be verified to satisfy the identity (\ref{eq:psp}).  This
actual reduction of the type IIB supersymmetry to $D=5$, $N=2$ verifies the
form of the superpotential assumed in \cite{Duffliustelle}.

Note that, at this stage, the scalars $\varphi$ and $f$ are still linear
combinations of the $E_0=8$ breathing and $E_0=6$ squashing mode on
$S^5$.  An $O(2)$ rotation, given in \cite{Bremer}, may be used to
disentangle these two modes.  From the $N=2$ point of view, the
breathing and squashing modes $(\varphi,f)$, along with the fermions
$(\lambda^{(5)},\chi^{(5)})$, belong to a massive vector representation
of $SU(2,2|1)$%
\footnote{Unitary highest weight representations of $SU(2,2|1)$ were
constructed in \cite{Flato:1984te,Dobrev:1985qv}, while those of
$SU(2,2|N/2)$ were investigated in
\cite{Bars:1983ep,Gunaydin:1988hb,Gunaydin:1999jc,Gunaydin:1998sw}.
See also \cite{ceresole,Freedman:1999gp}.}.
Denoting AdS$_5$ representations by $D(E_0,j_1,j_2;r)$ where $r$ is
the $U(1)$ charge, the content of this vector multiplet is given by
\begin{eqnarray}
\label{eq:bsm}
{\cal D}(6,0,0;0)&=&D(7,\ft12,\ft12;0)\nonumber\\
&&+D(6\ft12,\ft12,0;-1)+D(6\ft12,0,\ft12;1)+D(7\ft12,0,\ft12;-1)
+D(7\ft12,\ft12,0;1)\nonumber\\
&&+D(6,0,0;0)+D(7,0,0;-2)+D(7,0,0;2)+D(8,0,0;0).
\end{eqnarray}
The breathing mode and squashing mode scalars can be identified with
$D(8,0,0;0)$ and $D(6,0,0;0)$, respectively.
In addition to the fields we have considered, this indicates the
presence of a charged scalar, $D(7,0,0;-2)+D(7,0,0;2)$, and vector,
$D(7,\fft12,\fft12;0)$.  The latter presumably has its origin in
${\cal A}_{[1]}$.

The Kaluza-Klein spectrum for the round $S^5$ compactification of type
IIB theory was obtained in \cite{Gunaydin,Kim:1985ez}, and falls into
unitary representations of $SU(2,2|4)$.  Following the above procedure
of squashing the five-sphere, this $N=8$ supersymmetry may be
broken by decomposing $SU(2,2|4)\supset SU(2,2|1)\times SU(3)\times U(1)$
and truncating to the $SU(3)$ singlet sector \cite{Ferrara}.  For the
massless supergravity sector, this decomposition yields the $N=2$ gravity
multiplet coupled to a LH$+$RH chiral multiplet (which contains the type
IIB dilaton and Ramond-Ramond scalar):
\begin{eqnarray}
{\cal D}(3,\ft12,\ft12;0)&=&D(4,1,1;0)
+D(3\ft12,1,\ft12;-1)+D(3\ft12,\ft12,1;1)+D(3,\ft12,\ft12;0),\nonumber\\
{\cal D}(3,0,0,2)&=&D(3\ft12,\ft12,0;1)+D(3,0,0;2)+D(4,0,0;0),\nonumber\\
{\cal D}(3,0,0,-2)&=&D(3\ft12,0,\ft12;-1)+D(3,0,0;-2)+D(4,0,0;0).
\end{eqnarray}
At the first Kaluza-Klein level, the $N=2$ truncation yields a semi-long
LH$+$RH massive gravitino multiplet:
\begin{eqnarray}
{\cal D}(4\ft12,0,\ft12,1)&=&D(5\ft12,\ft12,1;1)
+D(5,\ft12,\ft12;0)+D(5,0,1;2)+D(6,0,1,0)\nonumber\\
&&+D(4\ft12,0,\ft12,1)+D(5\ft12,0,\ft12,-1),\nonumber\\
{\cal D}(4\ft12,\ft12,0,-1)&=&D(5\ft12,1,\ft12;-1)
+D(5,\ft12,\ft12;0)+D(5,1,0;-2)+D(6,1,0,0)\nonumber\\
&&+D(4\ft12,\ft12,0,-1)+D(5\ft12,\ft12,0,1).
\end{eqnarray}
Finally, at the second Kaluza-Klein level, the truncation yields
precisely the $N=2$ breathing/squashing multiplet given in
(\ref{eq:bsm}).  These decompositions agree with the bosonic sector of
the five-dimensional Lagrangian obtained in \cite{Bremer}.


\section{Reduction of $D=11$ Supergravity to four dimensions}

Our second example of the supersymmetry of breathing mode compactifications
concerns the reduction of eleven dimensional supergravity on $S^7$
\cite{kksugra}.  In this
case, we start with the bosonic fields, $\hat G_{MN}$ and $\hat
F_{[4]}=dA_{[3]}$, with Lagrangian
\begin{equation}
\label{eq:11lag}
e^{-1}{\cal L}_{11}=\hat R-\fft1{2\cdot4!}F_{[4]}^2
- \fft16 e^{-1} *(F_{[4]}\wedge F_{[4]}\wedge A_{[3]}).
\end{equation}
For a bosonic background, the $D=11$ gravitino supersymmetry transformation
is given by \cite{Cremmer}
\begin{equation}
\label{eq:11gravi}
\delta\hat\psi_{M} = \left[\hat\nabla_M
-\frac{1}{288}(\Gamma_M{}^{PQRS}-8\delta_M^P\Gamma^{QRS})F_{PQRS}
\right]\hat\epsilon.
\end{equation}
As in the type IIB scenario, we first consider the case of a round $S^7$,
followed by the turning on of a squashing mode introduced by writing $S^7$
as a $U(1)$ bundle over $CP^3$.

\subsection{The breathing mode}

Following the general sphere reduction of \cite{Bremer}, for the round $S^7$
we choose the standard ansatz
\bea
\label{eq:kka4}
ds_{11}^2&=&e^{2\alpha\varphi}ds_4^2+e^{2\beta\varphi}ds^2(S^7)\nonumber\\
F_{[4]}&=&ce^{6\alpha\varphi}\epsilon_{[4]},
\eea
where $\epsilon_{[4]}$ is the volume form in the $D=4$ spacetime.  For the
case at hand, $\alpha$ and $\beta$ take on the values
\begin{equation}
\alpha=\fft{\sqrt{7}}6,\qquad\beta=-\fft27\alpha.
\end{equation}
The resulting four-dimensional bosonic Lagrangian reads \cite{Bremer}
\begin{equation}
e^{-1}{\cal L}_4 = R-\ft12(\partial\varphi)^2-V(\phi),
\end{equation}
where
\begin{equation}
\label{eq:4pot}
V=\half c^2e^{6\alpha\varphi}-R_{7}e^{\frac{18}{7}\alpha\varphi}.
\end{equation}
This potential has an AdS$_4$ minimum at
\begin{equation}
e^{\fft{24}7\alpha\varphi_*}=\fft{6R_7}{7c^2},
\end{equation}
where $R_7$ is the Ricci scalar of $S^7$.

The reduction of (\ref{eq:11gravi}) is now straightforward, and follows
the procedure developed in the previous section.  Thus we omit the details,
and only point out some salient features of the reduction.  For spinors, we
start with a natural $4+7$ split of the Dirac matrices:
\be
\label{eq:47spl}
\Gamma^{M}=(\gamma^{\mu}\otimes 1,\gamma^{5}\otimes\tilde\gamma^{a}),
\ee
where $\gamma^{5}=i\gamma^{0}\gamma^{1}\gamma^{2}\gamma^{3}$ is the
spacetime chirality matrix (and squares to $+1$).  Using this decomposition,
the spacetime and sphere components of (\ref{eq:11gravi}) become
\begin{eqnarray}
\delta\hat\psi_{\mu}&=&\left[\nabla_\mu
+\fft\alpha2\gamma_{\mu}{}^\nu\partial_\nu\varphi
-\fft{i}{6}c e^{3\alpha\varphi}\gamma_{\mu}\gamma^{5}\right]\hat\epsilon,
\nonumber\\
\delta\hat\psi_a&=&\left[\tilde\nabla_a
-\fft\alpha7 e^{-\fft97\alpha\varphi}\gamma^{5}(\gamma\cdot\partial\varphi)
\otimes \tilde\gamma_a +\fft{i}{12}c e^{\fft{12}7\alpha\varphi}
\tilde\gamma_a\right ]\hat\epsilon.
\end{eqnarray}
Defining the shifted four-dimensional quantities as
\begin{eqnarray}
\varepsilon^{(4)}_i\otimes\eta^i&=&e^{-\fft12\alpha\varphi}\hat\epsilon,
\nonumber\\
\lambda^{(4)}_i\otimes\eta^i&=&-\fft1{2\alpha}e^{\fft{11}{14}\alpha\varphi}
\gamma^5\tilde\gamma^a\hat\psi_a
,\nonumber\\
\psi_{i\,\mu}^{(4)}\otimes\eta^i&=&e^{-\fft12\alpha\varphi}\hat\psi_\mu
-\alpha\gamma_\mu\lambda^{(4)}_i\otimes\eta^i,
\end{eqnarray}
and making use of the Killing spinor equation on the sphere,
(\ref{eq:snkse}), we finally obtain the four-dimensional supersymmetry
variations
\begin{eqnarray}
\delta\psi_{i\,\mu}^{(4)}&=&\left[\nabla_\mu+\fft{i}8\left(ce^{3\alpha\varphi}
\mp14\sqrt{\fft{R_4}{42}}e^{\fft97\alpha\varphi}\right)\gamma_\mu\gamma^5
\right]\varepsilon_i^{(4)},\nonumber\\
\delta\lambda_i^{(4)}&=&\fft12\left[\gamma\cdot\partial\varphi
-\fft{7i}{12\alpha}\left(ce^{3\alpha\varphi}\mp6\sqrt{\fft{R_4}{42}}
e^{\fft97\alpha\varphi}\right)\gamma^5\right]\varepsilon_i^{(4)}.
\end{eqnarray}
The index $i$ runs from 1 to 8, and labels the $D=4$, $N=8$ supersymmetry
arising from having eight Killing spinors on $S^7$.  Furthermore, as before,
the choice of sign depends on the orientation of the sphere.

The factor $i\gamma^5$ may be rewritten for Majorana spinors in $D=4$, as
appropriate.  Regardless, there is a straightforward truncation to $N=2$
obtained by choosing a single Majorana spinor pair.  The resulting
transformations, written in a $D=4$, $N=2$ language, are
\begin{eqnarray}
\label{eq:4n2}
\delta\psi_\mu&=&\left[\nabla_\mu-\ft1{4\sqrt{2}}
W\gamma_\mu(i\gamma^5)\right]\varepsilon,\nonumber\\
\delta\lambda&=&\ft12
\left[\gamma\cdot\partial\varphi+\sqrt{2}\partial_\varphi W(i\gamma^5)\right]
\varepsilon,
\end{eqnarray}
where
\begin{equation}
W=-\fft1{\sqrt{2}}\left[ce^{3\alpha\varphi}\mp14\sqrt{\fft{R_7}{42}}
e^{\fft97\alpha\varphi}\right].
\end{equation}
This superpotential and the potential, (\ref{eq:4pot}), satisfy
the relation (\ref{eq:psp}).

\subsection{Introduction of a squashing mode}

Again since $S^7$ can be viewed as a $U(1)$ bundle over $CP^3$,
one can perform the reduction in two steps.  We first reduce
from eleven dimensions on a circle giving the type IIA theory in ten
dimensions.  Following this, we may proceed from ten down to four
dimensions on $CP^3$.  This approach to squashing $S^7$ has been extensively
studied in \cite{Nilsson,Duff:1997qz}.  However note that, as in the
$S^5$ case, we wish to retain states charged under the $U(1)$ fiber.  While
such states may be regarded as non-perturbative in a type IIA point of view
\cite{Duff:1997qz}, they naturally arise from eleven dimensions and complete
the supersymmetry of the compactification.

As seen in \cite{Nilsson,Duff:1997qz}, the $U(1)$ neutral sector has either
$N=6$ or $N=0$ supersymmetry in four dimensions, depending on the
orientation of $S^7$.  This is easily seen by considering the decomposition
of the spinor $8_s$ under $SO(8)\supset SU(4)\times U(1)$:
$8_s\to 6_0+1_2+1_{-2}$ for left squashing and $8_s\to 4_1+\overline{4}_{-1}$
for right squashing.  Thus in the round $S^7$ limit, complete $N=8$
supersymmetry requires the introduction of $U(1)$ charged spinors.
Furthermore, we are mainly interested in the truncation of the
breathing/squashing system to $N=2$, which corresponds to the $SU(4)$
singlet sector under the above decomposition.  We see that the left squashed
compactification has precisely the expected $SU(4)$ singlet supercharges
corresponding to $U(1)$ gauged $D=4$, $N=2$ supergravity.  Curiously, this
$N=2$ supersymmetry is in some sense complementary to the $N=6$
supersymmetry considered previously in the $U(1)$ neutral sector
\cite{Nilsson}.

There has been a long tradition, starting with Refs.~\cite{Campbell,Huq},
of reducing eleven dimensional supergravity on a circle in order to
obtain the type IIA theory.  The reduction proceeds with the metric ansatz
\be
ds_{11}^2=e^{2a\phi}ds_{10}{}^2+e^{2b\phi}(dz+{\cal A}_Mdx^M)^2,
\ee
where
\begin{equation}
a=-\frac{1}{12}, \qquad b=-8a.
\end{equation}
The resulting type IIA supergravity is described by the bosonic Lagrangian
\begin{equation}
e^{-1}{\cal L}_{10}=R-\ft12(\partial\phi)^2-\ft1{2\cdot2!}e^{\fft32\phi}
{\cal F}_{[2]}^2-\ft1{2\cdot3!}e^{-\phi}F_{[3]}^2-\ft1{2\cdot4!}e^{\fft12\phi}
F_{[4]}^2-\ft12*(F_{[4]}\wedge F_{[4]}\wedge A_{[2]}),
\end{equation}
where $F_{[4]}$ is now shifted, $F_{[4]}=dA_{[3]}-F_{[3]}\wedge
{\cal A}_{[1]}$.

Eleven-dimensional Dirac matrices may be given in terms of their
ten-dimensional counterparts by setting $\Gamma^{10}=\Gamma^{11}$ where
$\Gamma^{11}\equiv\Gamma^0\Gamma^1\cdots\Gamma^9$ is the $D=10$
chirality operator.  In this case, the resulting supersymmetry
transformations are \cite{Campbell,Huq}
\bea
\label{eq:10dsusy}
\delta\psi_M^{(10)}&=&\biggl[D_M + \fft18e^{-\fft34\phi}\Gamma_M\Gamma^{11}
\partial_z-{1\over 64} e^{\fft34\phi}
(\Gamma_M{}^{NP}-14\delta_M^N\Gamma^P){\cal F}_{NP}\Gamma^{11}
\nonumber\\
&&\qquad-\fft1{4\cdot4!}e^{-\fft12\phi}
(\Gamma_M{}^{NPQ}-9\delta_M^N\Gamma^{PQ}) F_{NPQ}\Gamma^{11}\nonumber\\
&&\qquad-\fft1{256}e^{\fft14\phi}
(\Gamma_M{}^{NPQR}-\fft{20}3\delta_M^N\Gamma^{PQR}) F_{NPQR}
\biggr]\epsilon^{(10)},\nonumber\\
\delta\lambda^{(10)}&=&\fft12\biggl[\Gamma\cdot\partial\phi
+3e^{-\fft34\phi}\Gamma^{11}\partial_z
-\fft38e^{\fft34\phi}{\cal F}_{[2]}\cdot\Gamma\Gamma^{11}\nonumber\\
&&\kern6em+\fft1{12}e^{-\fft12\phi}F_3\cdot\Gamma\Gamma^{11}
-\fft1{96}e^{\fft14\phi}F_4\cdot\Gamma\biggr]\epsilon^{(10)}.
\eea
The $D=10$ quantities are related to the original ones by
\begin{eqnarray}
\psi_M^{(10)}&=&e^{\fft1{24}\phi}(\hat\psi_M-{\cal A}_M\hat\psi_z)
+\ft1{12}\Gamma_M\lambda^{(10)},\nonumber\\
\lambda^{(10)}&=&\ft32e^{-\fft{17}{24}\phi}\Gamma^{11}\hat\psi_z,\nonumber\\
\epsilon^{(10)}&=&e^{\fft1{24}\phi}\hat\epsilon.
\end{eqnarray}
As in the squashing of $S^5$, we have retained momentum dependence in the
$z$ direction, corresponding to $U(1)$ charged spinors.  The $U(1)$
covariant derivative is given by $D_M=\nabla_M-{\cal A}_M\partial_z$.

For a reduction on $CP^3$, we choose the ansatz \cite{Duff:1997qz,Bremer}
\bea
\label{eq:kkacp3}
ds_{10}^2&=&e^{2\alpha\varphi}ds_4^2+e^{2\beta\varphi}ds^2(CP^3),\nonumber\\
F_{[4]}&=&ce^{-\fft12\phi+6\alpha\varphi}\epsilon_{(4)},\nonumber\\
F_{[3]}&=&0,\nonumber\\
{\cal F}_{[2]}&=&2mJ_{[2]}(CP^3),
\eea
with
\begin{equation}
\alpha=\fft{\sqrt{3}}4,\qquad\beta=-\fft13\alpha.
\end{equation}
Note that this ansatz sets to zero the $D=4$ field strengths originating
from $F_{[3]}$ and ${\cal F}_{[2]}$ (since our interest is only on the
breathing/squashing scalars).  The resulting $D=4$ Lagrangian has the form
\begin{equation}
e^{-1}{\cal L}_4=R-\ft12(\partial\phi)^2-\ft12(\partial\varphi)^2
-V(\phi,\varphi),
\end{equation}
where
\begin{equation}
V=6m^2e^{\fft32\phi+\fft{10}3\alpha\varphi}
+\fft12c^2e^{-\fft12\phi+6\alpha\varphi}-R_6e^{\fft83\alpha\varphi}.
\end{equation}
The AdS$_4$ minimum of this potential lies at
\begin{equation}
\label{eq:n8crit}
c^2=36m^2e^{2\phi_*-\fft83\alpha\varphi_*},\qquad
R_6=48m^2e^{\fft32\phi_*+\fft23\alpha\varphi_*},
\end{equation}
which is in fact the round $S^7$ vacuum.

For reduction of the supersymmetry variations, (\ref{eq:10dsusy}), we may
use the decomposition of the Dirac matrices, (\ref{eq:47spl}), specialized
now to the present $10=4+6$ split.  With this split, we find
$\Gamma^{11}=-\gamma^5\otimes\tilde\gamma^7$, where
$\tilde\gamma^7=i\tilde\gamma^1\tilde\gamma^2\cdots\tilde\gamma^6$.
Substituting the ansatz, (\ref{eq:kkacp3}), into (\ref{eq:10dsusy}), we find
\begin{eqnarray}
\label{eq:10var}
\delta\psi_\mu^{(10)}&=&\biggl[\nabla_\mu+\fft\alpha2\gamma_\mu{}^\nu
\partial_\nu\varphi-\fft{i}{32}\gamma_\mu\gamma^5\biggl(
4\left(\fft{R_6}{48m}\right)qe^{-\fft34\phi+\alpha\varphi}\tilde\gamma^7
\nonumber\\
&&\kern12em+m\tilde Qe^{\fft34\phi+\fft53\alpha\varphi}
+5ce^{-\fft14\phi+3\alpha\varphi}\biggr)\biggr]\epsilon^{(10)},\nonumber\\
\delta\psi_a^{(10)}&=&\biggl[\tilde D_a-\fft\alpha6e^{-\fft43\alpha\varphi}
\gamma^5\gamma\cdot\partial\varphi\tilde\gamma_a-\fft{i}8
\left(\fft{R_6}{48m}\right)qe^{-\fft34\phi-\fft13\alpha\varphi}
\tilde\gamma_a\tilde\gamma^7\nonumber\\
&&\qquad\qquad+\fft{m}{32}e^{\fft34\phi+\fft13\alpha\varphi}
(\tilde\gamma_a{}^{bc}-14\delta_a^b\tilde\gamma^c)J_{bc}\tilde\gamma^7
+\fft{3ic}{32}e^{-\fft14\phi+\fft53\alpha\varphi}\tilde\gamma_a\biggr]
\epsilon^{(10)},\nonumber\\
\delta\lambda^{(10)}&=&\fft12e^{-\alpha\varphi}\biggl[
\gamma\cdot\partial\phi-3i\gamma^5\biggl(\left(\fft{R_6}{48m}\right)
qe^{-\fft34\phi+\alpha\varphi}\tilde\gamma^7\nonumber\\
&&\kern12em+\fft{m}4\tilde Q
e^{\fft34\phi+\fft53\alpha\varphi}-\fft{c}{12}e^{-\fft14\phi+3\alpha
\varphi}\biggr)\biggr]\epsilon^{(10)}.
\end{eqnarray}
As in the case of $U(1)$ bundled over $CP^2$, we use the definition
\begin{equation}
\tilde Q=iJ_{[2]}\cdot\tilde\gamma \gamma^7.
\end{equation}
Furthermore, using the same argument as before, we have identified the
period along the circle direction, $z=z+2\pi L$, to be $L=48m/R_6$
\cite{Duff:1997qz}.  The $U(1)$ charge is then related to the Kaluza-Klein
momentum through $\partial_z=iq/L$.

The Killing spinors $\eta$ on the Hopf fibered $S^7$ may be obtained by
examination of the variations (\ref{eq:10var}) at the $N=8$ critical point
given by (\ref{eq:n8crit}).  Corresponding to the two possibilities
\begin{equation}
c=\pm6me^{\phi_*-\fft43\alpha\varphi_*},
\end{equation}
we find
\begin{eqnarray}
\delta\lambda^{(10)}&=&-\fft{3i}8e^{\fft13\alpha\varphi_*}\gamma^5
\sqrt{\fft{R_6}{48}}(4q\tilde\gamma^7+\tilde Q\mp2)\epsilon^{(10)},\nonumber\\
\delta(\psi_a^{(10)}-\ft1{12}e^{-\fft13\alpha\varphi_*}\gamma^5\tilde\gamma_a
\lambda^{(10)})&=&\biggl[\tilde D_a-\fft{i}2\sqrt{\fft{R_6}{48}}
(\mp\tilde\gamma_a-iJ_{ab}\tilde\gamma^b\tilde\gamma^7)\biggr]\epsilon^{(10)}.
\end{eqnarray}
As a result, Killing spinors must satisfy
\begin{equation}
\label{eq:qQcond7}
(4q\tilde\gamma^7+\tilde Q\mp2)\eta=0
\end{equation}
[compare with (\ref{eq:qQcond})].  For $CP^3$, $\tilde Q$ has eigenvalues
2 (six times) and $-6$ (twice).  For either the top or bottom choice of sign
in (\ref{eq:qQcond7}), corresponding to left- or right-squashing respectively,
we may find the appropriate $U(1)$ charge $q$ such that the Killing spinor
condition is satisfied.  The result is shown in Table~\ref{tbl:qvals}, and
agrees with the charge assignments obtained from the spinor decompositions
$8_s\to 6_0+1_2+1_{-2}$ (left-squashing) and $8_s\to4_1+\overline{4}_{-1}$
(right-squashing).  Furthermore, it is also straightforward to check
the integrability of the gravitino variation on $CP^3$, following the
procedure given previously for the case of $CP^2$.

\begin{table}[t]
\begin{center}
\begin{tabular}{|c|c|c|c|}
\hline
$\tilde Q$&$\tilde\gamma^7$&$q$ (left-squashing)&$q$ (right-squashing)\\
\hline
$-6$&$-1$&$-2$&$-1$\\
$-6$&$\hphantom{-}1$&$\hphantom{-}2$&$\hphantom{-}1$\\
$\hphantom{-}2$&$-1$&$\hphantom{-}0$&$\hphantom{-}1$\\
$\hphantom{-}2$&$-1$&$\hphantom{-}0$&$\hphantom{-}1$\\
$\hphantom{-}2$&$-1$&$\hphantom{-}0$&$\hphantom{-}1$\\
$\hphantom{-}2$&$\hphantom{-}1$&$\hphantom{-}0$&$-1$\\
$\hphantom{-}2$&$\hphantom{-}1$&$\hphantom{-}0$&$-1$\\
$\hphantom{-}2$&$\hphantom{-}1$&$\hphantom{-}0$&$-1$\\
\hline
\end{tabular}
\end{center}
\label{tbl:qvals}
\caption{The $\tilde Q$ and $\tilde\gamma^7$ eigenvalues and $U(1)$ charges
$q$ for eight Killing spinors on the $U(1)$ over $CP^3$ fibered $S^7$.}
\end{table}

Using the effective Killing spinor equation,
\be
\left(\tilde\gamma^{a}\tilde D_a+\fft{i}2\sqrt{\frac{R_6}{48}}(\tilde Q\pm6)
\right)\eta=0,
\ee
obtained from (\ref{eq:qQcond7}) to eliminate $\tilde D_a$ from
(\ref{eq:10var}), we finally arrive at the set of $D=4$ supersymmetry
variations
\begin{eqnarray}
\delta\lambda^{(4)}&=&\fft12\biggl[\gamma\cdot\partial\phi
-\fft{3i}4\gamma^5\biggl(-\left(\fft{R_6}{48m}\right)(\tilde Q\mp2)
e^{-\fft34\phi+\alpha\varphi}+m\tilde Qe^{\fft34\phi+\fft53\alpha\varphi}
-\fft{c}3e^{-\fft14+3\alpha\varphi}\biggr)\biggr]\varepsilon^{(4)},\nonumber\\
\delta\chi^{(4)}&=&\fft12\biggl[\gamma\cdot\partial\varphi
+\fft{i}{16\alpha}\gamma^5\biggl(8\sqrt{\fft{R_6}{48}}(\tilde Q\pm6)
e^{\fft43\alpha\varphi}-3\left(\fft{R_6}{48m}\right)(\tilde Q\mp2)
e^{-\fft34\phi+\alpha\varphi}\nonumber\\
&&\kern16em-5m\tilde Qe^{\fft34\phi+\fft53\alpha\varphi}
-9ce^{-\fft14+3\alpha\varphi}\biggr)\biggr]\varepsilon^{(4)},\nonumber\\
\delta\psi_\mu^{(4)}&=&\biggl[\nabla_\mu-\fft{i}8\gamma_\mu\gamma^5
\biggl(2\sqrt{\fft{R_6}{48}}(\tilde Q\pm6)
e^{\fft43\alpha\varphi}-\left(\fft{R_6}{48m}\right)(\tilde Q\mp2)
e^{-\fft34\phi+\alpha\varphi}\nonumber\\
&&\kern16em-m\tilde Qe^{\fft34\phi+\fft53\alpha\varphi}
-ce^{-\fft14+3\alpha\varphi}\biggr)\biggr]\varepsilon^{(4)}.
\end{eqnarray}
The four-dimensional spinors are related to the original ten-dimensional
ones through
\begin{eqnarray}
\varepsilon^{(4)}\otimes\eta&=&e^{-\fft12\alpha\varphi}\epsilon^{(10)},
\nonumber\\
\lambda^{(4)}\otimes\eta&=&e^{\fft12\alpha\varphi}\lambda^{(10)},
\nonumber\\
\chi^{(4)}\otimes\eta&=&-\ft1{2\alpha}e^{\fft56\alpha\varphi}\gamma^5
\tilde\gamma_a\psi_a^{(10)},
\nonumber\\
\psi^{(4)}_\mu\otimes\eta&=&e^{\fft12\alpha\varphi}\psi_\mu^{(10)}
-\alpha\gamma_\mu\chi^{(4)}\otimes\eta.
\end{eqnarray}

Once again, these variations may be written in an explicit $D=4$, $N=2$
manner, given by (\ref{eq:4n2}), where the superpotential has the form
\begin{equation}
W=-\fft1{\sqrt{2}}\biggl[-2\sqrt{\fft{R_6}{48}}(\tilde Q\pm6)
e^{\fft43\alpha\varphi}+\left(\fft{R_6}{48m}\right)(\tilde Q\mp2)
e^{-\fft34\phi+\alpha\varphi}+m\tilde Qe^{\fft34\phi+\fft53\alpha\varphi}
+ce^{-\fft14+3\alpha\varphi}\biggr].
\end{equation}
The only case where supersymmetry survives truncation to the $SU(4)$ singlet
sector is for left-squashing (the top sign), in which case the two
$\tilde Q=-6$ states may be combined into a single $U(1)$ charged Dirac
spinor.  The resulting $N=2$ superpotential has the form
\begin{equation}
W=-\fft1{\sqrt{2}}\biggl[ce^{-\fft14+3\alpha\varphi}
-6me^{\fft34\phi+\fft53\alpha\varphi}
-\fft{R_6}{6m}e^{-\fft34\phi+\alpha\varphi}\biggr],
\end{equation}
and satisfies the $N=2$ relation (\ref{eq:psp}).

Once again, it is instructive to examine the decomposition of $D=4$,
$N=8$ (for the round $S^7$) to $D=4$, $N=2$ supersymmetry.  In this
case, $N=2$ is preserved only for left-squashing, where $8_s\to
6_0+1_2+1_{-2}$ under $SO(8)\supset SU(4)\times U(1)$.  Truncating to
$SU(4)$ singlets, the massless ($n=0$) Kaluza-Klein sector yields the
pure $N=2$ supergravity multiplet
\begin{equation}
{\cal D}(2,1;0)=D(3,2;0)+D(\ft52,\ft32;-1)+D(\ft52,\ft32;1)+D(2,1;0).
\end{equation}
In contrast to the five-dimensional case, no states at the $n=1$ level
survive the truncation.  The breathing mode finally shows up at the
second Kaluza-Klein level, where the truncation to $N=2$ yields a
massive vector
\begin{eqnarray}
{\cal D}(4,0;0)&=&D(5,1;0)\\
&&+D(\ft92,\ft12;-1)+D(\ft92,\ft12;1)
+D(\ft{11}2,\ft12;-1)+D(\ft{11}2,\ft12;1)\nonumber\\
&&+D(4,0;0)+D(5,0;0)+D(5,0;-2)+D(5,0;2)+D(6,0;0).\nonumber
\end{eqnarray}
As before, the representations are given in terms of $E_0$, spin and
$U(1)$ charge.  The breathing mode is identified as the $E_0=6$ scalar,
while the squashing mode is the $E_0=4$ scalar.  The remaining neutral
(axionic) scalar may be identified with the three-form field strength
$F_{[3]}$, which was set to zero in (\ref{eq:kkacp3}) but could as well
have been retained \cite{Bremer}.


\section{Reduction of $D=11$ Supergravity to seven Dimensions}

The last case we consider is the reduction of eleven dimensional
supergravity on $S^4$.  Since the even sphere cannot be written as a Hopf
fibration, we cannot play the same trick to turn on a single squashing mode.
Thus we focus only on the round $S^4$.

For this case, the reduction ansatz is \cite{Bremer}
\bea
ds_{11}^2&=&e^{2\alpha\varphi}ds_7^2+e^{2\beta\varphi}ds^2(S^4),\nonumber\\
{\hat F}_{[4]}&=&F_{[4]}+6m\epsilon_{[4]}(S^4),
\eea
with
\be
\alpha=\fft2{3\sqrt{10}},\qquad\beta=-\fft54\alpha,
\ee
whereupon the bosonic Lagrangian, (\ref{eq:11lag}), reduces to
\begin{equation}
e^{-1}{\cal L}_7=R-\ft12(\partial\varphi)^2-\ft1{2\cdot4!}e^{-6\alpha\varphi}
F_{[4]}^2-3m e^{-1}*(F_{[4]}\wedge A_{[3]})-V(\varphi),
\end{equation}
where
\begin{equation}
V=18m^2e^{12\alpha\varphi}-R_4e^{\fft92\alpha\varphi}.
\end{equation}
Note that, following \cite{Bremer}, we have retained a possible four-form
field strength, $F_{[4]}$, in seven dimensions.  The potential has a minimum
at
\begin{equation}
e^{\fft{15}2\alpha\varphi_*}=\fft{R_4}{48m^2},
\end{equation}
giving rise to an AdS$_7$ vacuum, provided $F_{[4]}=0$.

In a $7+4$ split, the Dirac matrices decompose as
\be
\Gamma^{M}=(\gamma^{\mu}\otimes \tilde\gamma^{5},1\otimes\tilde\gamma^{a})
\ee
where $\tilde\gamma^{5}=\tilde\gamma^{1}\tilde\gamma^{2}\tilde\gamma^{3}
\tilde\gamma^{4}$ is the chirality operator on $S^4$.  In this case the
$D=11$ gravitino supersymmetry transformation, (\ref{eq:11gravi}), reduces to
\begin{eqnarray}
\delta\hat\psi_{\mu}&=&\left[\nabla_\mu+\fft\alpha2
\gamma_{\mu}{}^\nu\partial_\nu\varphi
-\fft{m}2 e^{6\alpha\varphi}\gamma_{\mu}
-\frac{1}{12\cdot4!}e^{-3\alpha\varphi}(\gamma_{\mu}{}^{\nu\rho\lambda\sigma}
-8\delta_\mu^\nu\gamma^{\rho\lambda\sigma})\tilde\gamma^5
F_{\nu\rho\lambda\sigma}\right]\hat\epsilon,\nonumber\\
\delta\hat\psi_{a}&=&\left[\tilde\nabla_a
-\fft{5\alpha}8 e^{-\fft94\alpha\varphi} (\gamma\cdot\partial\varphi)
\otimes \tilde\gamma_{a}\tilde\gamma^5
+m e^{\fft{15}4\alpha\varphi}\tilde\gamma_{a}\tilde\gamma^5
-\frac{1}{12\cdot4!}e^{-\fft{21}4\alpha\varphi}
F_{[4]}\cdot\gamma \otimes \tilde\gamma_{a}\right]\hat\epsilon.\nonumber\\
\end{eqnarray}
For this case, the appropriate Killing spinors on $S^4$ satisfy
\begin{equation}
\left[\tilde\nabla_a\pm\fft12\tilde\gamma_a\tilde\gamma^5\sqrt{\fft{R_4}{12}}
\right]\eta=0,
\end{equation}
yielding four Killing spinors $\eta^i$, $i=1,2,3,4$ and leading to the
$D=7$, $N=4$ transformations
\begin{eqnarray}
\delta\psi_{i\,\mu}^{(7)}&=& \Biggl[ \nabla_{\mu}
+\fft1{10}\Biggl(3me^{6\alpha\varphi}
\mp4\sqrt{\frac{R_4}{12}}e^{\fft94\alpha\varphi}\Biggr)
\gamma_{\mu}\nonumber\\
&&\kern9.4em-\frac{1}{480}e^{-3\alpha\varphi}
(3\gamma_\mu{}^{\nu\rho\lambda\sigma}
-8\delta_\mu^\nu\gamma^{\rho\lambda\sigma})\tilde\gamma^5
F_{\nu\rho\lambda\sigma}
\Biggr] \varepsilon^{(7)}_i,\nonumber\\
\delta\lambda_i^{(7)}&=&\fft12\Biggl[\gamma\cdot\partial\varphi
-\fft4{5\alpha}\Biggl(2me^{6\alpha\varphi}\mp\sqrt{\fft{R_4}{12}}
e^{\fft94\alpha\varphi}\Biggr)
+\fft1{180\alpha}e^{-3\alpha\varphi}F_{[4]}\cdot\gamma\tilde\gamma^5
\Biggr]\varepsilon^{(7)}_i.
\end{eqnarray}
Here the seven-dimensional quantities are defined as
\bea
\varepsilon_i^{(7)}\otimes\eta^i&=&e^{-\fft12\alpha\varphi}\hat\epsilon,
\nonumber\\
\lambda_i^{(7)}\otimes\eta^i&=&-\fft1{5\alpha}e^{\fft74\alpha\varphi}
\tilde\gamma^5\tilde\gamma^a\hat\psi_a,
\nonumber\\
\psi_{i\,\mu}^{(7)}&=&e^{-\fft12\alpha\varphi}\hat\psi_\mu-\alpha\gamma_\mu
\lambda_i^{(7)}\otimes\eta^i.
\eea

The above equations can be written more suggestively as
\bea
\delta\psi_{\mu}&=&\left[\nabla_{\mu}-\ft1{10\sqrt{2}}W\gamma_\mu
-\ft1{480}e^{-3\alpha\varphi}(3\gamma_\mu{}^{\nu\rho\lambda\sigma}
-8\delta_\mu^\nu\gamma^{\rho\lambda\sigma})\tilde\gamma^5
F_{\nu\rho\lambda\sigma}\right]\varepsilon,\nonumber\\
\delta\lambda&=&\ft12\left[\gamma\cdot\varphi+\sqrt{2}\partial_\varphi W
+\ft1{180\alpha}e^{-3\alpha\varphi}F_{[4]}\cdot\gamma\tilde\gamma^5
\right]\varepsilon,
\eea
where the superpotential is identified as
\be
W=-\sqrt{2}\left[3me^{6\alpha\varphi}
\mp4\sqrt{\frac{R_4}{12}}e^{\frac{9}{4}\alpha\varphi}\right].
\ee
Note that this satisfies the identity (\ref{eq:psp}) as expected.


\section{Discussion}

In the above, we have considered the supersymmetry of breathing mode
compactifications for the three cases: $D=10$ on $S^5$, $D=11$ on $S^7$ and
$D=11$ on $S^4$.  In all cases, the breathing mode is a singlet under the
$R$ symmetry, despite the fact that it lies in the massive Kaluza-Klein
spectrum.  For this reason, inclusion of the breathing mode in itself
is allowed in a consistent truncation of the full Kaluza-Klein spectrum.
However, for the resulting theory to be (maximally) supersymmetric,
the superpartners
to the breathing mode must also be retained.  Presumably once these
non-singlet superpartners are included, this would no longer be a consistent
truncation unless the entire Kaluza-Klein tower is brought in as well.
Thus it may not be entirely appropriate to regard this breathing mode
compactification as solely a lower dimensional supergravity theory coupled
to the breathing mode supermultiplet.  While consideration of the higher
dimensional equations of motion ensures the validity of the above
reduction ansatse, the resulting theory of the form (\ref{eq:n=2lag})
and (\ref{eq:n=2}) is necessarily incomplete.  The complete structure of
the theory, and especially its supersymmetry, is perhaps more naturally
seen in the original higher dimensional form.

On the other hand, for the squashed sphere compactifications, one may
consistently truncate to the $SU(n+1)$ singlet sector of the full $R$
symmetry group.  Since this procedure already removes many states in
the Kaluza-Klein tower, it suggests that the $N=2$ truncations of the
$D=5$ and $D=4$ theories may admit a further truncation yielding only
the breathing/squashing multiplet coupled to the massless supergravity
multiplet.  If this were in fact the case, it would provide an interesting
example of a consistent truncation to a massive supergravity theory
where only a portion of the Kaluza-Klein tower survives.

Finally, note that recent investigation of the brane-world has shown that
the `kinked' Randall-Sundrum geometry is only compatible with supersymmetry
provided the superpotential $W$ changes sign when passing through the brane
\cite{Kallosh:2000tj,BeCvII,%
Altendorfer:2000rr,Falkowski:2000er,Bergshoeff:2000zn}.
Restoring the gauge coupling constant $g$, this corresponds in the
five-dimensional point of view to $g\to -g$ on opposite sides of the
brane \cite{Falkowski:2000er,Bergshoeff:2000zn}.  This cannot be realized
in a strictly $D=5$, $N=2$ point of view, but however is expected from the
type IIB compactification, where it corresponds to a reversal of the
orientation of $S^5$.  This orientation flip corresponds to making the
opposite choice of sign in the Killing spinor equation on the sphere,
(\ref{eq:snkse}).  Furthermore, the five-form flux changes sign,
$m\to -m$, so that the superpotential, (\ref{eq:5spot}) indeed flips
sign as expected.  This is also the case for the superpotential on the
squashed $S^5$, (\ref{eq:5sqsp}).

However, by changing the sign in the Killing spinor equation,
(\ref{eq:snkse}), this orientation reversal joins opposite sets of
Killing spinors on both sides of the Randall-Sundrum brane.  This
potential difficulty with supersymmetry is even more pronounced for the
$S^7$ case where left- and right-squashing yield rather different
realizations of the Killing spinors (cf.~Table~\ref{tbl:qvals}).  Of
course, the kinked brane-world is singular at the location of the brane.
So perhaps it is not surprising to see this behavior of the Killing
spinors upon orientation reversal.  It remains to be seen what effect
this has on a complete understanding of the supersymmetry of the
brane-world.

\section*{Acknowledgements}
We would like to thank M.J.~Duff for useful discussions, especially
in regards to supersymmetry and Killing spinors on the squashed spheres.
JTL wishes to thank the Center for Advanced Mathematical Sciences (CAMS)
at the American University of Beirut, where part of this work was
performed.  This research was supported in part by DOE Grant
DE-FG02-95ER40899 Task G.


\end{document}